\documentclass{ws-mpla}

\usepackage{epsfig}

\newcommand{\lsim}{
\mathrel{\hbox{\rlap{\hbox{\lower4pt\hbox{$\sim$}}}\hbox{$<$}}}}

\newcommand{\gsim}{
\mathrel{\hbox{\rlap{\hbox{\lower4pt\hbox{$\sim$}}}\hbox{$>$}}}}

\begin{document}

\markboth{Robert Fleischer}
{Exploring CP Violation through B Decays}

%
\catchline{}{}{}{}{}
%


\thispagestyle{empty}

\begin{flushright}
CERN-TH/2003-115\\
hep-ph/0305267
\end{flushright}

\vspace{3cm}
\begin{center}
\boldmath
\large\bf Exploring CP Violation through B Decays 
\unboldmath
\end{center}

\vspace{1.2cm}
\begin{center}
Robert Fleischer\\[0.1cm]
{\sl Theory Division, CERN, CH-1211 Geneva 23, Switzerland}
\end{center}

\vspace{1.7cm}

\begin{center}
{\bf Abstract}
\end{center}

{\small
\vspace{0.2cm}\noindent
The $B$-meson system provides many strategies to perform stringent
tests of the Standard-Model description of CP violation. In this
brief review, we discuss implications of the currently available
$B$-factory data on the angles $\alpha$, $\beta$ and $\gamma$ of 
the unitarity triangle, emphasize the importance of $B_s$ studies 
at hadronic $B$ experiments, and discuss new, theoretically clean
strategies to determine $\gamma$.}

\vspace{1.7cm}

\begin{center}
{\sl Invited brief review for Modern Physics Letters {\bf A}}
\end{center}

\vfill
\noindent
CERN-TH/2003-115\\
May 2003

\newpage
\thispagestyle{empty}
\vbox{}
\newpage
 
\setcounter{page}{1}


\title{EXPLORING CP VIOLATION THROUGH B DECAYS}

\author{\footnotesize ROBERT FLEISCHER}

\address{Theory Division, CERN\\ 
CH-1211 Geneva 23, Switzerland\\
Robert.Fleischer@cern.ch}

\maketitle

\pub{Received (Day Month Year)}{Revised (Day Month Year)}

\begin{abstract}
The $B$-meson system provides many strategies to perform stringent
tests of the Standard-Model description of CP violation. In this
brief review, we discuss implications of the currently available
$B$-factory data on the angles $\alpha$, $\beta$ and $\gamma$ of 
the unitarity triangle, emphasize the importance of $B_s$ studies 
at hadronic $B$ experiments, and discuss new, theoretically clean
strategies to determine $\gamma$.
\keywords{CP violation; unitarity triangle; non-leptonic $B$ decays.}
\end{abstract}

\ccode{PACS Nos.: 12.15.Hh, 13.25.Hw, 11.30.Er.}

\section{Introduction}\label{sec:intro}
The discovery of CP violation through the observation of 
$K_{\rm L}\to\pi^+\pi^-$ decays in 1964 came as a big 
surprise.\cite{CP-discovery} This particular kind of CP violation,
which is referred to as ``indirect'' CP violation and is
described by a complex quantity $\varepsilon_K$, originates from 
the fact that the $K_{\rm L}$ mass eigenstate is not a pure eigenstate 
of the CP operator with eigenvalue $-1$, but one that receives a tiny 
admixture of the CP eigenstate with eigenvalue $+1$. Another milestone in
the exploration of CP violation through neutral kaon decays came in 
1999, when also ``direct'' CP violation, i.e.\ CP-violating effects 
arising directly at the amplitude level, could be established by
the NA48 (CERN) and KTeV (FNAL) collaborations through a measurement 
of a non-vanishing value of 
$\mbox{Re}(\varepsilon_K'/\varepsilon_K)$.\cite{eps-prime} Unfortunately, 
this observable does not provide a stringent test of the Standard-Model 
description of CP violation, unless significant theoretical progress 
concerning the relevant hadronic matrix elements can be 
made.\cite{bertolini,buras-KAON} 

One of the phenomenologically most exciting topics in this decade is 
the exploration of decays of $B$ mesons, which offer various powerful 
tests of the CP-violating sector of the Standard Model (SM) and 
provide, moreover, valuable insights into hadron dynamics.\cite{RF-Phys-Rep} 
The experimental stage is now governed by the asymmetric $e^+e^-$ $B$ 
factories operating at the $\Upsilon(4S)$ resonance, with their detectors 
BaBar (SLAC) and Belle (KEK). These experiments have already established 
CP violation in the $B$-meson system in 2001, which is the beginning of a 
new era in the exploration of CP violation.\cite{BaBar-CP-obs,Belle-CP-obs}
Many interesting strategies can now be confronted with data.\cite{BABAR-BOOK} 
In the near future, also run II of the Tevatron is expected to contribute 
significantly to this programme, providing -- among other things -- first 
access to $B_s$-meson decays.\cite{TEV-BOOK} In the era of the LHC, these 
decay modes can then be fully exploited,\cite{LHC-BOOK} in particular at 
LHCb (CERN) and BTeV (FNAL). 

\begin{figure}
\begin{tabular}{lr}
   \epsfysize=3.6cm
   \epsffile{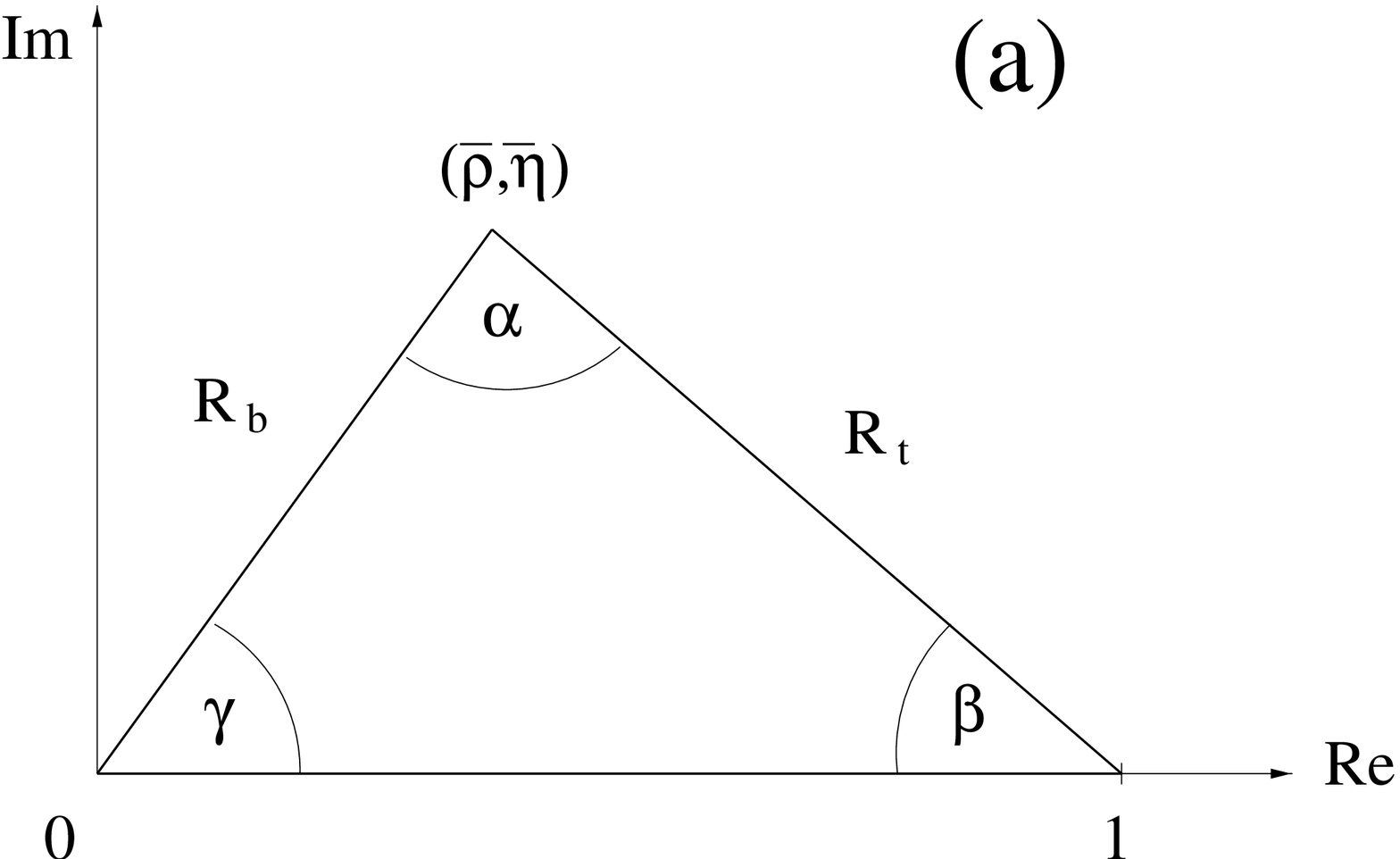}
&
   \epsfysize=3.6cm
   \epsffile{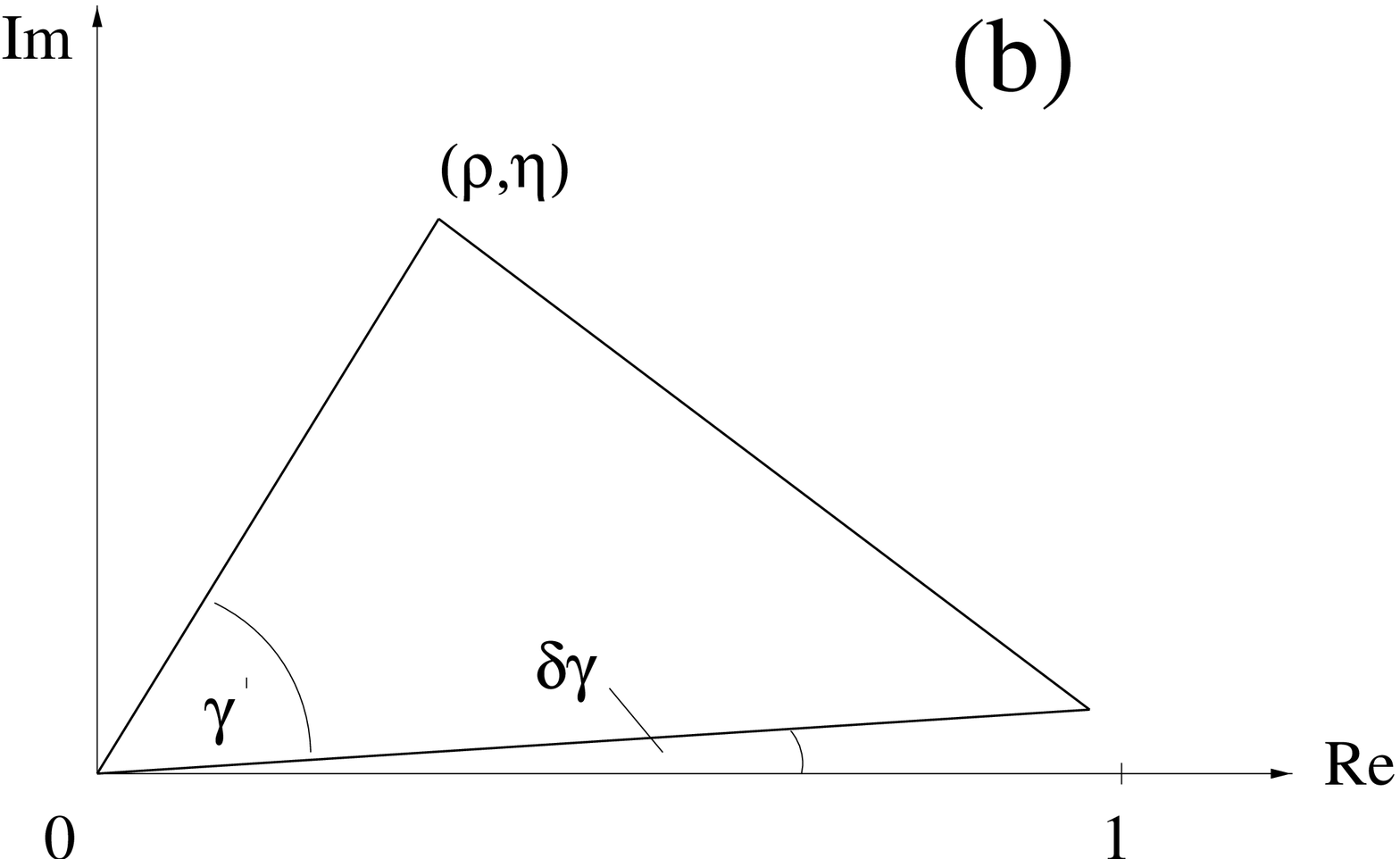}
\end{tabular}
\caption[]{The two non-squashed unitarity triangles of the CKM matrix,
where (a) and (b) correspond to 
$V_{ud}V_{ub}^\ast+V_{cd}V_{cb}^\ast+V_{td}V_{tb}^\ast=0$ and
$V_{ub}^\ast V_{tb}+V_{us}^\ast V_{ts}+V_{ud}^\ast V_{td}=0$, 
respectively.}\label{fig:UT}
\end{figure}

Within the framework of the SM of electroweak interactions, CP violation 
can be accommodated through a complex phase 
in the Cabibbo--Kobayashi--Maskawa (CKM) matrix, as pointed out by 
Kobayashi and Maskawa in 1973.\cite{KM} Since the CKM matrix is
unitary, we obtain six orthogonality relations, which can be represented
in the complex plane as six triangles. Applying the Wolfenstein 
parametrization,\cite{wolf} we find that there are only two 
phenomenologically interesting triangles, where all three sides are of
the same order in $\lambda\equiv|V_{us}|=0.22$. If we divide all sides
by $A\equiv|V_{cb}|/\lambda^2$, we obtain the triangles illustrated in
Fig.~\ref{fig:UT}, where $\overline{\rho}$ and $\overline{\eta}$ are 
related to $\rho$ and $\eta$ through 
$\overline{\rho}\equiv\left(1-\lambda^2/2\right)\rho$ and
$\overline{\eta}\equiv\left(1-\lambda^2/2\right)\eta$, 
respectively,\cite{blo} 
and $\delta\gamma\equiv\gamma-\gamma'=\lambda^2\eta$. Whenever we 
refer to a unitarity triangle (UT) in the following discussion, 
we mean the one shown in Fig.\ \ref{fig:UT} (a).

The main goal is now to overconstrain the UT as much
as possible, with the hope to encounter discrepancies, which may shed 
light on new physics (NP). To this end, we may, on the one hand, obtain 
indirect information on the UT angles $\alpha$, $\beta$ and $\gamma$ 
through measurements of the UT sides $R_b$ and $R_t$: the former can be 
fixed through exclusive and inclusive semileptonic $B$ decays caused by
$b\to u\ell\overline{\nu}_{\ell}, c\ell\overline{\nu}_{\ell}$ quark-level
transitions, whereas $R_t$ can be determined, within the SM, with the
help of $B^0_q$--$\overline{B^0_q}$ mixing ($q\in\{d,s\}$). Finally, 
using the SM interpretation of the indirect CP violation in the neutral kaon 
system, which is measured by $\varepsilon_K$, we may fix a hyperbola in the
$\overline{\rho}$--$\overline{\eta}$ plane. Many different strategies to 
deal with the corresponding theoretical and experimental uncertainties 
can be found in the literature; a detailed discussion of these approaches 
is beyond the scope of this brief review, and was recently given 
in Ref.\ \refcite{CKM-Proc}. Let us here just list  typical ranges for 
$\alpha$, $\beta$ and $\gamma$ that follow from such ``CKM fits'':
\begin{equation}\label{UT-Fit-ranges}
70^\circ\lsim\alpha\lsim 130^\circ, \quad
20^\circ\lsim\beta\lsim 30^\circ, \quad
50^\circ\lsim\gamma\lsim 70^\circ.
\end{equation}
If we measure, on the other hand, CP-violating effects in $B$-meson decays,
we may obtain {\it direct} information on the UT angles $\alpha$, $\beta$ 
and $\gamma$, as well as on $\delta\gamma$. This topic will be the focus
of the following considerations. 

In Section~\ref{sec:strat}, we shall classify the main strategies to
explore CP violation through $B$ decays. The implications of
the currently available $B$-factory data for various benchmark modes
will then be discussed in Section~\ref{sec:bench}, whereas we shall
focus on the $B_s$-meson system -- the ``El Dorado'' for
$B$-physics studies at hadron colliders -- in Section~\ref{sec:Bs}.
In Section~\ref{sec:gam-clean}, we turn to new, theoretically clean 
strategies to extract $\gamma$, and finish in Section~\ref{sec:concl} 
by summarizing our conclusions and giving a brief outlook.

\section{Classification of the Main Strategies}\label{sec:strat}
In the exploration of CP violation through $B$ decays, non-leptonic 
transitions play the key r\^ole, since CP asymmetries may there be
induced through certain interference effects. In particular, if
the decay receives contributions from two amplitudes with non-trivial
weak and strong phase differences, we obtain {\it direct} CP violation.
Unfortunately, hadronic matrix elements of local four-quark operators,
which are very hard to estimate in a reliable manner, enter the calculation 
of such CP asymmetries, thereby precluding a clean extraction of the weak
phase differences, which are related to the angles of the UT and are 
usually given by $\gamma$. Nevertheless, interesting progress could 
recently be made in this very challenging direction through the development 
of the QCD factorization\cite{BBNS1}$^{\mbox{--}}$\cite{fact-recent} and  
perturbative hard-scattering (PQCD)\cite{PQCD} formalisms, QCD light-cone 
sum-rule approaches,\cite{sum-rules} and soft collinear effective
theory (SCET),\cite{SCET} as reviewed in detail in Ref.\ \refcite{li-ref}.

In order to extract solid information on the angles of the UT from 
CP-violating asymmetries, the theoretical input on hadronic matrix 
elements should be reduced as much as possible. Such strategies, allowing 
in particular the determination of $\gamma$, are provided by fortunate 
cases, where we may eliminate the hadronic matrix elements through relations 
between different decay amplitudes: we distinguish between exact relations, 
involving pure tree-diagram-like decays of the kind 
$B\to DK$\cite{gw}$^{\mbox{--}}$\,\cite{ADS} or $B_c\to D D_s$,\cite{FW} 
and relations, which follow from the flavour symmetries of strong 
interactions, involving $B_{(s)}\to \pi\pi, \pi K, KK$ 
decays.\cite{GRL}$^{\mbox{--}}$\cite{FlMa2} 

If we employ decays of neutral $B_d$ or $B_s$ mesons, another avenue 
to deal with the problems arising from hadronic matrix elements 
emerges. It is offered by a new kind of CP violation, which is
referred to as {\it mixing-induced} CP violation, and is due to 
interference effects between $B^0_q$--$\overline{B^0_q}$ 
($q\in\{d,s\}$) mixing and decay processes. In the time-dependent
rate asymmetry
\begin{eqnarray}
\lefteqn{\frac{\Gamma(B^0_q(t)\to f)-
\Gamma(\overline{B^0_q}(t)\to f)}{\Gamma(B^0_q(t)\to f)+
\Gamma(\overline{B^0_q}(t)\to f)}}\nonumber\\
&&=\left[\frac{{\cal A}_{\rm CP}^{\rm dir}(B_q\to f)\,\cos(\Delta M_q t)+
{\cal A}_{\rm CP}^{\rm mix}(B_q\to f)\,\sin(\Delta 
M_q t)}{\cosh(\Delta\Gamma_qt/2)-{\cal A}_{\rm 
\Delta\Gamma}(B_q\to f)\,\sinh(\Delta\Gamma_qt/2)}\right],\label{ACP-time}
\end{eqnarray}
where $\Delta M_q\equiv M^{(q)}_{\rm H}-M^{(q)}_{\rm L}$ and 
$\Delta\Gamma_q\equiv\Gamma^{(q)}_{\rm H}-\Gamma^{(q)}_{\rm L}$ are 
the mass and decay widths differences of the $B_q$ mass eigenstates 
(``heavy'' and ``light''), respectively, and $({\cal CP})|f\rangle=
\pm|f\rangle$, this road shows up in the form of the coefficient 
of the $\sin(\Delta M_q t)$ term, whereas the one of $\cos(\Delta M_q t)$ 
measures the direct CP-violating effects discussed above. If the decay 
$B_q\to f$ is dominated by a single CKM amplitude, the corresponding 
hadronic matrix element cancels in ${\cal A}_{\rm CP}^{\rm mix}(B_q\to f)$. 
This observable is then simply given by $\pm\sin(\phi_q-\phi_f)$, where 
$\phi_f$ and $\phi_q$ are the weak $B_q\to f$ decay and 
$B^0_q$--$\overline{B^0_q}$ mixing phases, respectively.\cite{RF-Phys-Rep}
Within the SM, the former phases are induced by the CKM matrix 
elements entering the $B_q\to f$ decay amplitude, whereas the latter 
are related to the famous box diagrams mediating $B^0_q$--$\overline{B^0_q}$
mixing and are given as follows:
\begin{equation}\label{phiq-SM}
\phi_q = 2 \mbox{arg}(V_{tq}^\ast V_{tb})=\left\{\begin{array}{cl}
+2\beta={\cal O}(50^\circ) & \mbox{($q=d$)}\\
-2\delta\gamma={\cal O}(-2^\circ) & 
\mbox{($q=s$).}
\end{array}\right.
\end{equation}
Mixing-induced CP violation plays a key r\^ole for several benchmark 
modes, and is also a powerful ingredient to complement analyses of $B$
decays, which are related to one another through amplitude relations 
of the kind discussed above.

\section{Status of Benchmark Modes}\label{sec:bench}
\subsection{$B\to J/\psi K$}\label{ssec:BpsiK}
One of the most famous $B$ decays, the ``gold-plated'' mode 
$B_d^0\to J/\psi\,K_{\rm S}$ to extract $\sin2\beta$,\cite{bisa}  
originates from $\overline{b}\to\overline{c}c\overline{s}$ quark-level 
processes. If we look at the corresponding Feynman diagrams arising
within the SM, we observe that it receives contributions both from tree 
and from penguin topologies, and that the decay amplitude takes the
following form:
\begin{equation}\label{BdpsiK-ampl2}
A(B_d^0\to J/\psi K_{\rm S})\propto\left[1+\lambda^2 a e^{i\theta}
e^{i\gamma}\right],
\end{equation}
where the hadronic parameter $a e^{i\theta}$ measures, sloppily speaking, 
the ratio of the penguin to tree contributions.\cite{RF-BdsPsiK} Since this 
parameter enters in a doubly Cabibbo-suppressed way, and is na\"\i vely 
expected to be of ${\cal O}(\overline{\lambda})$, where $\overline{\lambda}=
{\cal O}(\lambda)={\cal O}(0.2)$ is a ``generic'' expansion 
parameter,\cite{FM-BpsiK} we eventually arrive at  
\begin{eqnarray}
{\cal A}_{\rm CP}^{\rm dir}(B_d\to J/\psi K_{\rm S})&=&0+
{\cal O}(\overline{\lambda}^3)\label{BpsiK-CP-dir}\\
{\cal A}_{\rm CP}^{\rm mix}(B_d\to J/\psi K_{\rm S})&=&-\sin\phi_d+
{\cal O}(\overline{\lambda}^3).\label{BpsiK-CP-mix}
\end{eqnarray}
In 2001, the 
$B_d^0\to J/\psi K_{\rm S}$ decay and similar modes led to the
observation of CP violation in the $B$-meson system by the BaBar and Belle
collaborations.\cite{BaBar-CP-obs,Belle-CP-obs} The present status of 
$\sin\phi_d\stackrel{\rm SM}{=}\sin2\beta$ is given as follows:
\begin{equation}
\sin2\beta=\left\{\begin{array}{ll}
0.741\pm 0.067  \pm0.033  &
\mbox{(BaBar\cite{Babar-s2b-02})}\\
0.719\pm 0.074  \pm0.035  &
\mbox{(Belle\cite{Belle-s2b-02}),}
\end{array}\right.
\end{equation}
yielding the world average 
\begin{equation}\label{s2b-average}
\sin2\beta=0.734\pm0.054, 
\end{equation}
which agrees well with the results of the ``CKM fits'' of the UT summarized
in (\ref{UT-Fit-ranges}), implying $0.6\lsim\sin2\beta\lsim0.9$. 

In the LHC era, the experimental accuracy of the measurement of $\sin2\beta$ 
may be increased by one order of magnitude.\cite{LHC-BOOK} Such a tremendous 
experimental accuracy would then require deeper insights into the 
${\cal O}(\overline{\lambda}^3)$ corrections affecting (\ref{BpsiK-CP-mix}), 
which are caused by penguin effects. A possibility to control them 
is provided by the $B_s\to J/\psi K_{\rm S}$ channel.\cite{RF-BdsPsiK} 
Moreover, as can be seen in (\ref{BpsiK-CP-dir}), also direct CP violation 
in $B_d\to J/\psi K_{\rm S}$ allows us to probe these effects. So far, 
there are no experimental indications for a non-vanishing value of 
${\cal A}_{\rm CP}^{\rm dir}(B_d\to J/\psi K_{\rm S})$.

The agreement between the world average (\ref{s2b-average}) and the results 
of the ``CKM fits'' is striking. However, it should not be forgotten that
NP may nevertheless hide in 
${\cal A}_{\rm CP}^{\rm mix}(B_d\to J/\psi K_{\rm S})$. 
The point is that the key quantity is actually the $B^0_d$--$\overline{B^0_d}$
mixing phase $\phi_d$ itself; we obtain the twofold solution
\begin{equation}\label{phid-det}
\phi_d=\left(47^{+5}_{-4}\right)^\circ \, \lor \,
\left(133^{+4}_{-5}\right)^\circ,
\end{equation}
where the former value is in perfect agreement with
$40^\circ\lsim2\beta\stackrel{\rm SM}{=}\phi_d\lsim60^\circ$, which is implied 
by the ``CKM fits'', whereas the latter would correspond to NP. The two 
solutions can obviously be distinguished through a measurement of the sign 
of $\cos\phi_d$. To accomplish this important task, several strategies were
proposed.\cite{ambig} Unfortunately, their practical implementations 
are rather challenging. One of the most accessible approaches employs 
the time-dependent angular distribution of the 
$B_d\to J/\psi[\to\ell^+\ell^-] K^\ast[\to\pi^0K_{\rm S}]$ decay products, 
allowing us to extract $\mbox{sgn}(\cos\phi_d)$, if we fix the sign of a 
hadronic parameter $\cos\delta_f$, which involves a strong phase $\delta_f$, 
through factorization.\cite{DDF2,DFN} This analysis is already 
in progress at the $B$ factories.\cite{itoh} For hadron colliders, the
$B_d\to J/\psi\rho^0$, $B_s\to J/\psi\phi$ system is very interesting to 
probe the sign of $\cos\phi_d$.\cite{RF-ang} As we will see in 
Subsection~\ref{ssec:BsKK}, there is an interesting connection between
the two solutions for $\phi_d$ in (\ref{phid-det}) and constraints on 
$\gamma$, which is offered through CP violation in $B_d\to\pi^+\pi^-$ 
decays.\cite{FlMa2}

The preferred mechanism for NP to enter the CP-violating effects of
the ``gold-plated'' $B_d\to J/\psi K_{\rm S}$ channel is through 
$B^0_d$--$\overline{B^0_d}$ mixing.\cite{GNW} However, NP may, 
in principle, also enter at the $B\to J/\psi K$ amplitude 
level.\cite{growo} Estimates borrowed from effective field theory 
suggest that these effects are at most ${\cal O}(10\%)$ for a generic 
NP scale $\Lambda_{\rm NP}$ in the TeV regime; in order to obtain the 
whole picture, a set of appropriate observables can be introduced, 
employing $B_d\to J/\psi K_{\rm S}$ and its charged counterpart 
$B^\pm\to J/\psi K^\pm$.\cite{FM-BpsiK} These observables are already 
severely constrained through the $B$-factory data, and do not signal 
any deviation from the SM.

\subsection{$B\to\phi K$}
Decays of the kind $B\to \phi K$, which originate from
$\overline{b}\to\overline{s}s\overline{s}$ quark-level transitions,
provide another important testing ground for the SM description of
CP violation. These modes are governed by QCD penguins,\cite{BphiK-old} 
but also their EW penguin contributions are sizeable.\cite{RF-EWP,DH-PhiK}
Since such penguin topologies are absent at the tree level in the SM,
$B\to\phi K$ decays represent a sensitive probe to search for NP effects. 
Within the SM, we obtain the following 
relations:\cite{growo,RF-rev}$^{\mbox{--}}$\cite{FM-BphiK}
\begin{eqnarray}
{\cal A}_{\rm CP}^{\rm dir}(B_d\to \phi K_{\rm S})&=&0+
{\cal O}(\overline{\lambda}^2)\\
{\cal A}_{\rm CP}^{\rm mix}(B_d\to \phi K_{\rm S})&=&
{\cal A}_{\rm CP}^{\rm mix}(B_d\to J/\psi K_{\rm S})+
{\cal O}(\overline{\lambda}^2).
\end{eqnarray}
As in the case of the $B\to J/\psi K$ system,\cite{FM-BpsiK} a combined 
analysis of $B_d\to \phi K_{\rm S}$ and its charged counterpart
$B^\pm \to \phi K^\pm$ should be performed in order to obtain the whole 
picture.\cite{FM-BphiK} There is also the possibility of an unfortunate 
case, where NP cannot be distinguished from the SM.\cite{RF-Phys-Rep,FM-BphiK}

The present experimental status of CP violation in $B_d\to\phi K_{\rm S}$
is given as follows:
\begin{equation}\label{aCP-Bd-phiK-dir}
{\cal A}_{\rm CP}^{\rm dir}(B_d\to \phi K_{\rm S})=
\left\{\begin{array}{ll}
-0.80\pm0.38\pm0.12 &\mbox{(BaBar\cite{HdM})}\\
+0.56\pm0.41\pm0.16 &\mbox{(Belle\cite{Belle-BphiK})}
\end{array}\right.
\end{equation}
\begin{equation}
{\cal A}_{\rm CP}^{\rm mix}(B_d\to \phi K_{\rm S})=
\left\{\begin{array}{ll}
+0.18\pm0.51\pm0.07 &\mbox{(BaBar\cite{HdM})}\\
+0.73\pm0.64\pm0.22 &\mbox{(Belle\cite{Belle-BphiK}).}
\end{array}\right.
\end{equation}
Since we have, on the other hand, 
${\cal A}_{\rm CP}^{\rm mix}(B_d\to J/\psi K_{\rm S})=-0.734 \pm 0.054$, 
we arrive at a puzzling situation, which has already stimulated many
speculations about NP effects in $B_d\to\phi K_{\rm S}$.\cite{BPhiK-NP} 
However, because of the large experimental uncertainties and 
the recently reported results for the direct CP asymmetries in 
(\ref{aCP-Bd-phiK-dir}), it is probably too early to get too excited 
by the possibility of having large NP contributions to the 
$B_d\to\phi K_{\rm S}$ decay amplitude. Moreover, a recent BaBar analysis 
of direct CP violation in $B^\pm\to\phi K^\pm$ and $B\to\phi K^\ast$ 
transitions does not signal any effect.\cite{BaBar-Bphi-K-dir} It will be 
very interesting to observe the evolution of the $B$-factory data, also on 
$B_d\to \eta' K_{\rm S}$ and other related modes.

\subsection{$B\to\pi\pi$}\label{ssec:Bpipi}
The $B_d^0\to\pi^+\pi^-$ channel is another prominent $B$-meson
transition, originating from $\overline{b}\to\overline{u}u\overline{d}$ 
quark-level processes. In the SM, we may write
\begin{equation}\label{Bpipi-ampl}
A(B_d^0\to\pi^+\pi^-)\propto\left[e^{i\gamma}-de^{i\theta}\right],
\end{equation}
where the CP-conserving strong parameter $d e^{i\theta}$ measures the 
ratio of the penguin to tree contributions.\cite{RF-BsKK} 
In contrast to the $B_d^0\to J/\psi K_{\rm S}$ amplitude 
(\ref{BdpsiK-ampl2}), this parameter 
does {\it not} enter (\ref{Bpipi-ampl}) in a doubly Cabibbo-suppressed 
way, thereby leading to the well-known ``penguin problem'' in 
$B_d\to\pi^+\pi^-$. If we had negligible penguin contributions, i.e.\ 
$d=0$, the corresponding CP-violating observables were simply given as 
follows:
\begin{equation}
{\cal A}_{\rm CP}^{\rm dir}(B_d\to\pi^+\pi^-)=0, \quad
{\cal A}_{\rm CP}^{\rm mix}(B_d\to\pi^+\pi^-)=\sin(\phi_d+2\gamma)
\stackrel{\rm SM}{=}-\sin 2\alpha,
\end{equation}
where we have used the SM expression $\phi_d=2\beta$ and the 
unitarity relation $2\beta+2\gamma=2\pi-2\alpha$ in the last identity.
We observe that actually $\phi_d$ and $\gamma$ enter directly 
${\cal A}_{\rm CP}^{\rm mix}(B_d\to\pi^+\pi^-)$, and not $\alpha$. 
Consequently, since $\phi_d$ can be fixed straightforwardly through 
$B_d\to J/\psi K_{\rm S}$, we may use $B_d\to\pi^+\pi^-$ to probe 
$\gamma$.\cite{FlMa2,RF-BsKK,RF-Bpipi} Measurements of the CP-violating 
$B_d\to\pi^+\pi^-$ observables are already available:
\begin{equation}\label{Adir-exp}
{\cal A}_{\rm CP}^{\rm dir}(B_d\to\pi^+\pi^-)=\left\{
\begin{array}{ll}
-0.30\pm0.25\pm0.04 & \mbox{(BaBar\cite{BaBar-Bpipi})}\\
-0.77\pm0.27\pm0.08 & \mbox{(Belle\cite{Belle-Bpipi})}
\end{array}
\right.
\end{equation}
\begin{equation}\label{Amix-exp}
{\cal A}_{\rm CP}^{\rm mix}(B_d\to\pi^+\pi^-)=\left\{
\begin{array}{ll}
-0.02\pm0.34\pm0.05& \mbox{(BaBar\cite{BaBar-Bpipi})}\\
+1.23\pm0.41 ^{+0.07}_{-0.08} & \mbox{(Belle\cite{Belle-Bpipi}).}
\end{array}
\right.
\end{equation}
The BaBar and Belle results are unfortunately not fully consistent with
each other. If we form, nevertheless, the weighted averages of 
(\ref{Adir-exp}) and (\ref{Amix-exp}), applying the rules of the Particle 
Data Group (PDG), we obtain 
\begin{eqnarray}
{\cal A}_{\rm CP}^{\rm dir}(B_d\to\pi^+\pi^-)&=&-0.51\pm0.19 \,\, (0.23)
\label{Bpipi-CP-averages}\\
{\cal A}_{\rm CP}^{\rm mix}(B_d\to\pi^+\pi^-)&=&+0.49\pm0.27 \,\, (0.61),
\label{Bpipi-CP-averages2}
\end{eqnarray}
where the errors in brackets are those increased by the PDG scaling-factor 
procedure.\cite{PDG} Direct CP violation at this level would require large 
penguin contributions with large CP-conserving strong phases, which are
not suggested by the QCD factorization approach, pointing towards
${\cal A}_{\rm CP}^{\rm dir}(B_d\to\pi^+\pi^-)\sim +0.1.$\cite{BBNS3}
In addition to (\ref{Bpipi-CP-averages}), a significant impact of penguins 
on $B_d\to\pi^+\pi^-$ is also indicated by data on the 
$B\to\pi K,\pi\pi$ branching ratios,\cite{FlMa2,RF-Bpipi} as well as by 
theoretical considerations.\cite{BBNS3,PQCD-appl} Consequently, it is 
already evident that we {\it must} take the penguin contributions to 
$B_d\to\pi^+\pi^-$ into account in order to extract information on
the UT from the corresponding CP asymmetries. Many approaches 
to address this challenging problem were 
proposed.\cite{BBNS3,FlMa2,Bpipi-strategies} We shall return to this 
issue in Subsection~\ref{ssec:BsKK}, focusing on an approach to complement
$B_d\to\pi^+\pi^-$ with the $B_s\to K^+K^-$ channel.\cite{RF-BsKK}

\subsection{$B\to\pi K$}\label{ssec:BpiK}
Decays of this kind originate from $\overline{b}\to\overline{d}d\overline{s},
\overline{u}u\overline{s}$ quark-level transitions, and may receive 
contributions both from penguin and from tree topologies, where the latter 
are associated with the UT angle $\gamma$. Because of the tiny value of
the CKM factor $|V_{us}V_{ub}^\ast/(V_{ts}V_{tb}^\ast)|\approx0.02$, 
$B\to\pi K$ modes are interestingly dominated by QCD penguins, despite the 
loop suppression of these topologies. As far as electroweak (EW) penguins 
are concerned, their effects are expected to be negligible in the case of the
$B^0_d\to\pi^-K^+$, $B^+\to\pi^+K^0$ system, as they contribute here only in 
colour-suppressed form. On the other hand, EW penguins may also contribute in 
colour-allowed form to $B^+\to\pi^0K^+$ and $B^0_d\to\pi^0K^0$, and are 
hence expected to be sizeable in these modes, i.e.\ of the same order of 
magnitude as the tree topologies.

Interference effects between tree and penguin amplitudes allow us
to probe $\gamma$, where we may eliminate hadronic matrix elements
with the help of the flavour symmetries of strong interactions. 
As a starting point, we may use an isospin relation, suggesting
the following $B\to\pi K$ combinations to determine $\gamma$: the ``mixed'' 
$B^\pm\to\pi^\pm K$, $B_d\to\pi^\mp K^\pm$ 
system,\cite{PAPIII}$^{\mbox{--}}$\cite{defan} the ``charged'' 
$B^\pm\to\pi^\pm K$, $B^\pm\to\pi^0K^\pm$ 
system,\cite{NR}$^{\mbox{--}}$\cite{BF-neutral1} and the ``neutral'' 
$B_d\to\pi^0 K$, $B_d\to\pi^\mp K^\pm$ system.\cite{BF-neutral1,BF-neutral2} 

All three $B\to\pi K$ systems can be described by the same set of 
formulae by just making straightforward replacements of 
variables.\cite{BF-neutral1} Let us first focus on the charged and neutral 
$B\to\pi K$ systems. In order to determine $\gamma$ and the corresponding
strong phases, we have to introduce appropriate CP-conserving and 
CP-violating observables, which are given as follows:
\begin{equation}\label{charged-obs}
\left\{\begin{array}{c}R_{\rm c}\\A_0^{\rm c}\end{array}\right\}
\equiv2\left[\frac{\mbox{BR}(B^+\to\pi^0K^+)\pm
\mbox{BR}(B^-\to\pi^0K^-)}{\mbox{BR}(B^+\to\pi^+K^0)+
\mbox{BR}(B^-\to\pi^-\overline{K^0})}\right]
\end{equation}
\begin{equation}\label{neutral-obs}
\hspace*{0.2truecm}
\left\{\begin{array}{c}R_{\rm n}\\A_0^{\rm n}\end{array}\right\}
\equiv\frac{1}{2}\left[\frac{\mbox{BR}(B^0_d\to\pi^-K^+)\pm
\mbox{BR}(\overline{B^0_d}\to\pi^+K^-)}{\mbox{BR}(B^0_d\to\pi^0K^0)+
\mbox{BR}(\overline{B^0_d}\to\pi^0\overline{K^0})}\right].
\end{equation}
For the parametrization of these observables, we employ the isospin relation 
mentioned above, and assume that certain rescattering effects are small, 
which is in accordance with the QCD factorization
picture.\cite{BBNS1}$^{\mbox{--}}$\cite{BBNS3} Anomalously large 
rescattering processes would be indicated by data on $B\to KK$ modes, 
which are already strongly constrained by the $B$ factories, and could 
in principle be included through more elaborate 
strategies,\cite{defan,neubert-BpiK,BF-neutral1} which are, however, 
beyond the scope of this brief review. Following these lines, we obtain
\begin{eqnarray}
R_{\rm c,n}&=&1-2r_{\rm c,n}\left(\cos\gamma-q\right)\cos\delta_{\rm c,n}
+\left(1-2q\cos\gamma+q^2\right)r_{\rm c,n}^2\\
A_0^{\rm c,n}&=&2r_{\rm c,n}\sin\delta_{\rm c,n}\sin\gamma,
\end{eqnarray}
where the parameters $r_{\rm c,n}$, $q$ and 
$\delta_{\rm c,n}$ have the following physical interpretation:
$r_{\rm c,n}$ is a measure for the ratio of tree to penguin 
topologies, and can be fixed through $SU(3)$ arguments and data on 
$B^\pm\to\pi^\pm\pi^0$ modes,\cite{GRL} yielding $r_{\rm c,n}\sim0.2$. On the 
other hand, $q$ describes the ratio of EW penguin to tree contributions, 
and can be determined through $SU(3)$ arguments, 
yielding $q\sim 0.7$.\cite{NR} Finally, $\delta_{\rm c,n}$ is the 
CP-conserving strong phase between the tree and penguin amplitudes.

We observe that $R_{\rm c,n}$ and $A_0^{\rm c,n}$ depend on only two 
``unknown'' parameters, $\delta_{\rm c,n}$ and $\gamma$. If we vary them 
within their allowed ranges, i.e.\ 
$-180^\circ\leq \delta_{\rm c,n}\leq+180^\circ$ 
and $0^\circ\leq \gamma \leq180^\circ$, we obtain an allowed region in 
the $R_{\rm c,n}$--$A_0^{\rm c,n}$ plane.\cite{FlMa1,FlMa2} Should the
measured values of $R_{\rm c,n}$ and $A_0^{\rm c,n}$ lie outside this 
region, we would immediately have a signal for NP. On the other hand, 
should the measurements fall into the allowed range, $\gamma$ and 
$\delta_{\rm c,n}$ could be extracted. In this case, $\gamma$
could be compared with the results of alternative ``direct'' strategies 
and the range implied by the ``CKM fits'', whereas $\delta_{\rm c,n}$
would provide valuable insights into hadron dynamics.

\begin{figure}
\vspace*{-0.0cm}
$$\hspace*{-1.cm}
\epsfysize=0.18\textheight
\epsfxsize=0.25\textheight
\epsffile{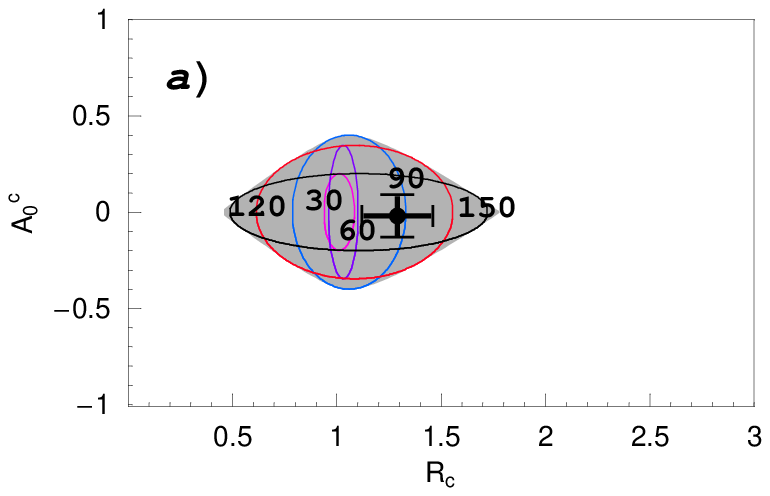} \hspace*{0.3cm}
\epsfysize=0.18\textheight
\epsfxsize=0.25\textheight
\epsffile{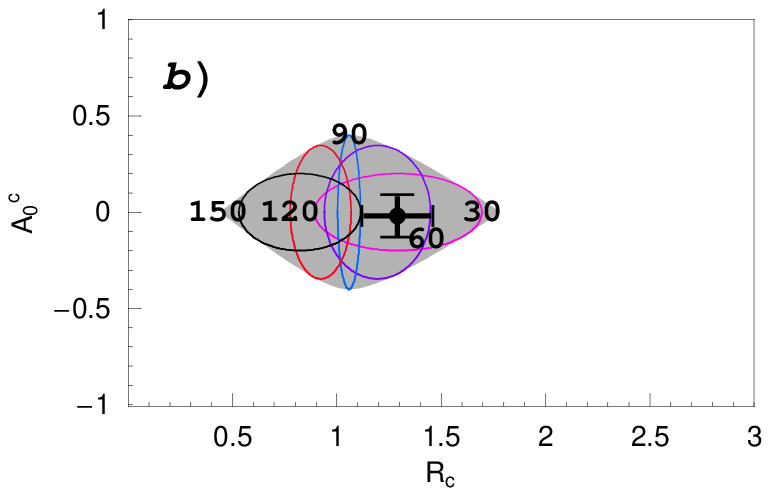}
$$
\vspace*{-0.9cm}
$$\hspace*{-1.cm}
\epsfysize=0.18\textheight
\epsfxsize=0.25\textheight
\epsffile{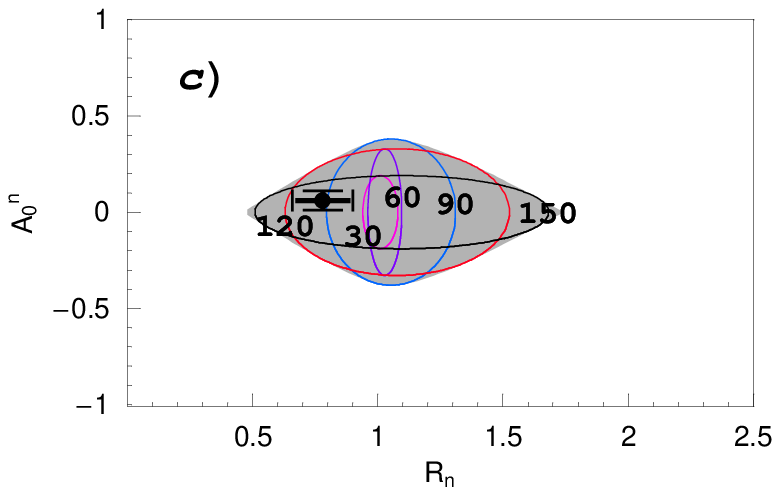} \hspace*{0.3cm}
\epsfysize=0.18\textheight
\epsfxsize=0.25\textheight
\epsffile{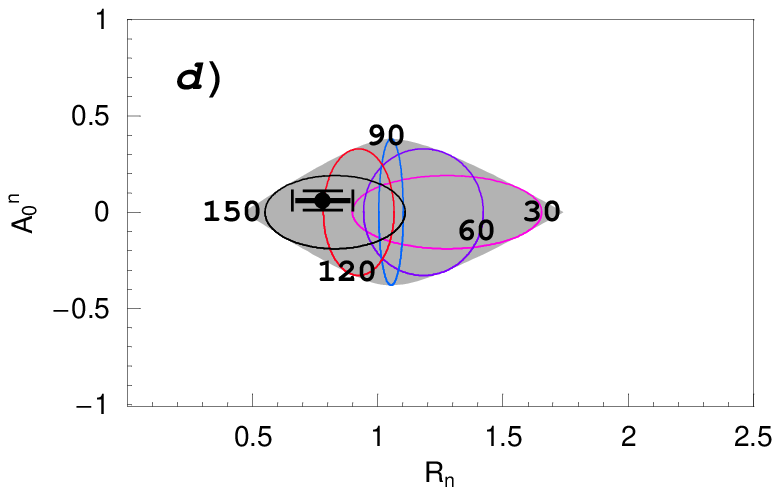}
$$
\caption[]{The allowed regions in the planes
$R_{\rm c}$--$A_0^{\rm c}$ ($r_{\rm c}=0.20$; (a), (b)) and
$R_{\rm n}$--$A_0^{\rm n}$ ($r_{\rm n}=0.19$; (c), (d)), 
for $q=0.68$: in (a) and (c), we show also the contours for fixed values of 
$\gamma$, whereas we give the curves arising for fixed values of 
$|\delta_{\rm c}|$ and $|\delta_{\rm n}|$ in (b) and (d), respectively.
}\label{fig:BpiK-OS}
\end{figure}

Following Ref.\ \refcite{FlMa2}, we show in Fig.~\ref{fig:BpiK-OS} the 
allowed regions in the $R_{\rm c,n}$--$A_0^{\rm c,n}$ planes, where the 
crosses represent the averages 
of the most recent $B$-factory data.\cite{CLEO-III,olsen} As can be 
read off from the contours in these figures, both the charged and the
neutral $B\to \pi K$ data favour $\gamma\gsim90^\circ$, which 
would be in conflict with (\ref{UT-Fit-ranges}). On the other hand, 
the charged modes point towards $|\delta_{\rm c}|\lsim90^\circ$ 
(factorization predicts $\delta_{\rm c}$ to be close to 
$0^\circ$\cite{BBNS3}), whereas the neutral decays seem to favour 
$|\delta_{\rm n}|\gsim90^\circ$. Since we do not expect that
$\delta_{\rm c}$ differs significantly from $\delta_{\rm n}$, we
arrive at a ``puzzling'' picture, which was already considered in
Ref.\ \refcite{BF-neutral2}. Unfortunately, the experimental 
uncertainties do not yet allow us to draw definite conclusions. 
As far as the mixed $B\to\pi K$ system is concerned, the data 
fall well into the SM region in observable space, and do not show
any ``anomalous'' behaviour at the moment.

\boldmath
\section{CP Violation in $B_s$ Decays}\label{sec:Bs}
\unboldmath
\subsection{General Remarks}
Since $\Upsilon(4S)$ states decay only to $B_{u,d}$ mesons, but not to 
$B_s$, these mesons are not accessible at the $e^+e^-$ $B$ factories 
operating at the $\Upsilon(4S)$ resonance. On the other hand, plenty of 
$B_s$ mesons are produced at hadron colliders, so that the $B_s$ 
system can be considered as the ``El Dorado'' for $B$-decay experiments 
at such machines.

The most exciting aspect of $B_s$ studies is clearly the exploration of 
CP violation. However, also the measurement of the mass difference 
$\Delta M_s$ would be a very important achievement, since this quantity 
nicely complements its $B_d$-meson counterpart $\Delta M_d$, thereby 
allowing a particularly interesting determination of 
the side $R_t$ of the UT. So far, only experimental lower bounds on 
$\Delta M_s$ are available, which can be converted into upper bounds on 
$R_t$, implying $\gamma\lsim 90^\circ$.\cite{CKM-Proc} In the near future, 
run II of the Tevatron should provide a measurement of $\Delta M_s$, 
thereby constraining the UT -- and in particular $\gamma$ -- in a much 
more stringent way. Interesting applications of $\Delta\Gamma_s$, which 
may be as large as ${\cal O}(10\%)$,\cite{BeLe} whereas $\Delta\Gamma_d$
is negligibly small, are extractions of $\gamma$ from ``untagged'' $B_s$ 
data samples, where we do not distinguish between initially, i.e.\ at 
time $t=0$, present $B^0_s$ or $\overline{B^0_s}$ 
mesons.\cite{Bs-untagged,FD2}

\boldmath
\subsection{$B_s\to J/\psi \phi$}\label{ssec:Bspsiphi}
\unboldmath
This decay is the $B_s$ counterpart of the 
``gold-plated'' mode $B_d\to J/\psi K_{\rm S}$, and originates from 
this channel if we replace the down spectator quark by a strange quark. 
Consequently, the phase structure of the $B_s\to J/\psi \phi$ decay 
amplitude is analogous to the one of (\ref{BdpsiK-ampl2}). However,
in contrast to $B_d\to J/\psi K_{\rm S}$, the final state of 
$B_s\to J/\psi\phi$ is an admixture of different CP eigenstates. In order
to disentangle them, we must measure the angular distribution of the 
$J/\psi\to \ell^+\ell^-$, $\phi\to K^+K^-$ decay products.\cite{DDLR} 
The corresponding CP-violating observables are governed by quantities
with the following structure:\cite{RF-Phys-Rep}
\begin{equation}\label{Bspsiphi-obs}
\xi^{(s)}_{\psi\phi}\,\propto\, e^{-i\phi_s}
\left[1-i\,\sin\gamma\times{\cal O}(\overline{\lambda}^3)\right],
\end{equation}
where the generic expansion parameter $\overline{\lambda}$ was introduced 
in Subsection~\ref{ssec:BpsiK}, and describes the impact of penguin 
contributions. Since we have $\phi_s=-2\lambda^2\eta={\cal O}(-0.03)$ in 
the SM, the extraction of $\phi_s^{\rm SM}$ from mixing-induced 
CP-violating effects arising in the time-dependent 
$B_s\to J/\psi[\to \ell^+\ell^-] \phi[\to K^+K^-]$ angular distribution 
is affected by generic hadronic uncertainties of ${\cal O}(10\%)$. 
These penguin effects, which may become an important issue for the 
LHC,\cite{LHC-BOOK} can be controlled through $B_d\to J/\psi\, \rho^0$, 
exhibiting also other interesting features.\cite{RF-ang} 

Because of the tiny mixing-induced CP asymmetries arising in 
$B_s\to J/\psi\phi$ within the SM, this mode is an interesting probe 
to search for NP contributions to $B^0_s$--$\overline{B^0_s}$
mixing.\cite{NiSi} A detailed discussion of ``smoking-gun'' signals of 
sizeable values of $\phi_s$ was given in Ref.\ \refcite{DFN}, where  
also methods to fix this phase {\it unambiguously} were proposed. The 
latter issue was also recently addressed in Ref.\ \refcite{RF-gam-eff-03}.

\boldmath
\subsection{Complementing $B_d\to \pi^+\pi^-$ through 
$B_s\to K^+K^-$}\label{ssec:BsKK}
\unboldmath
As we have seen in Subsection~\ref{ssec:Bpipi}, the extraction of
UT angles from the CP-violating $B_d\to\pi^+\pi^-$ asymmetries is
strongly affected by hadronic penguin effects. In order to deal with
this problem, the decay $B_s\to K^+K^-$ offers an interesting avenue
for $B$ experiments at hadron colliders.\cite{RF-BsKK} Within the
SM, we may write the CP asymmetries provided by these modes in the
following form:
\begin{equation}\label{Bpipi-obs}
{\cal A}_{\rm CP}^{\rm dir}(B_d\to\pi^+\pi^-)=
\mbox{fct}(d,\theta,\gamma), \quad
{\cal A}_{\rm CP}^{\rm mix}(B_d\to\pi^+\pi^-)=
\mbox{fct}(d,\theta,\gamma,\phi_d)
\end{equation}
\begin{equation}\label{BsKK-obs}
{\cal A}_{\rm CP}^{\rm dir}(B_s\to K^+K^-)=
\mbox{fct}(d',\theta',\gamma), \quad
{\cal A}_{\rm CP}^{\rm mix}(B_s\to K^+K^-)=
\mbox{fct}(d',\theta',\gamma,\phi_s),
\end{equation}
where the hadronic quantities $d'$ and $\theta'$ are the $B_s\to K^+K^-$
counterparts of the parameters introduced in (\ref{Bpipi-ampl}). If 
we take into account that $\phi_d$ and $\phi_s$ can straightforwardly 
be fixed separately, we may use the CP-violating asymmetries of the 
$B_d\to\pi^+\pi^-$ and $B_s\to K^+K^-$ modes to determine $d$ and $d'$ 
as functions of $\gamma$, respectively. This can be done in a 
{\it theoretically clean} manner, i.e.\ without using flavour-symmetry
or plausible dynamical assumptions. If we look at the corresponding 
Feynman diagrams, we observe that $B_d\to\pi^+\pi^-$ is related to 
$B_s\to K^+K^-$ through an interchange of all down and strange quarks. 
Because of this feature, the $U$-spin flavour symmetry of strong 
interactions implies
\begin{equation}\label{U-spin-rel}
d'=d, \quad \theta'=\theta. 
\end{equation}
If we now apply the former relation, we may determine $\gamma$, as well 
as the strong phases $\theta'$ and $\theta$, which provide a nice
consistency check of the latter $U$-spin relation.\cite{RF-BsKK} 
This strategy is also very promising from an experimental point of view: 
at Tevatron-II and the LHC, experimental accuracies for 
$\gamma$ of ${\cal O}(10^\circ)$ and ${\cal O}(1^\circ)$, 
respectively, are expected.\cite{TEV-BOOK,LHC-BOOK}  
For a collection of other $U$-spin 
strategies, see Refs.\ \refcite{RF-BdsPsiK,RF-ang,U-spin-strat}.

\begin{figure}[t]
$$\hspace*{-1.cm}
\epsfysize=0.2\textheight
\epsfxsize=0.3\textheight
\epsffile{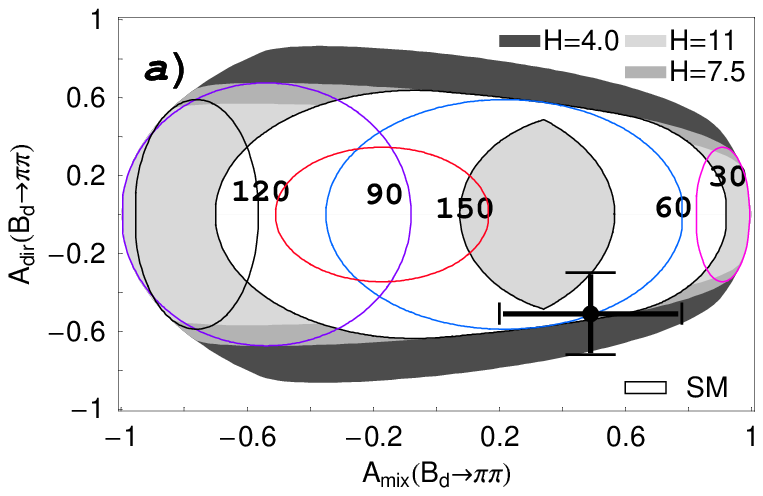} \hspace*{0.3cm}
\epsfysize=0.2\textheight
\epsfxsize=0.3\textheight
\epsffile{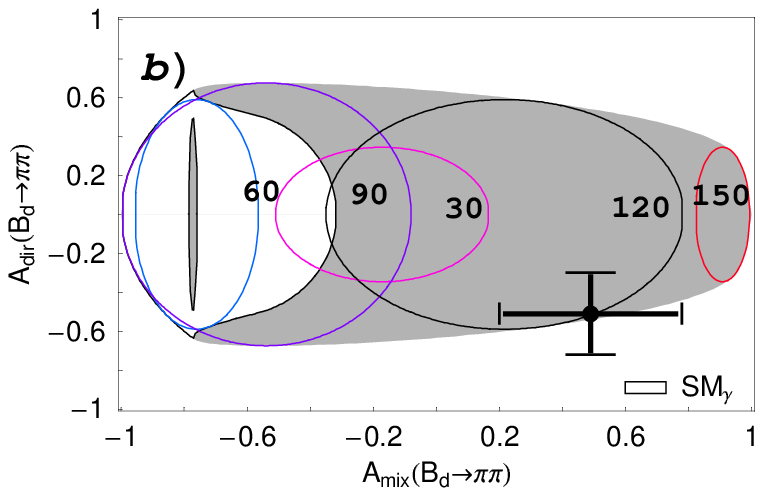}
$$
\caption[]{The allowed regions in the $B_d\to\pi^+\pi^-$ observable space,
where (a) was calculated for $\phi_d=47^\circ$ and various values of 
$H$, and (b) corresponds to $\phi_d=133^\circ$ and $H=7.5$. The SM 
regions appear if we restrict $\gamma$ to (\ref{UT-Fit-ranges}). 
Contours representing fixed values of $\gamma$ are also 
included.}\label{fig:AdAmpipi}
\end{figure}

Since $B_s\to K^+K^-$ is not accessible at the $e^+e^-$ $B$ factories 
operating at the $\Upsilon(4S)$ resonance, we may not yet implement this 
approach. However, $B_s\to K^+K^-$ is related to $B_d\to\pi^\mp K^\pm$ 
through an interchange of spectator quarks. Consequently, we may 
approximately replace $B_s\to K^+K^-$ through $B_d\to\pi^\mp K^\pm$ to 
deal with the penguin problem in $B_d\to\pi^+\pi^-$.\cite{RF-Bpipi}
To this end, the quantity
\begin{equation}\label{H-det}
H=\frac{1}{\epsilon}\left(\frac{f_K}{f_\pi}\right)^2
\left[\frac{\mbox{BR}(B_d\to\pi^+\pi^-)}{\mbox{BR}(B_d\to\pi^\mp K^\pm)}
\right]=
\left\{\begin{array}{ll}
7.4\pm2.5 & \mbox{(CLEO\cite{CLEO-III})}\\
7.8\pm1.2 & \mbox{(BaBar\cite{olsen})}\\
7.1\pm1.2 & \mbox{(Belle\cite{olsen}),}
\end{array}\right.
\end{equation}
where $\epsilon\equiv\lambda^2/(1-\lambda^2)$, is particularly useful.
Applying (\ref{U-spin-rel}), we may write
\begin{equation}\label{H-expr}
H=\mbox{fct}(d,\theta,\gamma).
\end{equation}
Consequently, if we complement (\ref{Bpipi-obs}) with (\ref{H-expr}), 
we have sufficient information to determine $\gamma$, $d$ and $\theta$. 
In particular, we may eliminate $d$ in 
${\cal A}_{\rm CP}^{\rm dir}(B_d\to\pi^+\pi^-)$ and 
${\cal A}_{\rm CP}^{\rm mix}(B_d\to\pi^+\pi^-)$, so that these observables 
then depend -- for a given value of $\phi_d$ -- only on $\gamma$ and 
the strong phase $\theta$. If we vary these parameters within 
their allowed ranges, we obtain an allowed region in the 
${\cal A}_{\rm CP}^{\rm dir}(B_d\to\pi^+\pi^-)$--${\cal A}_{\rm CP}^{\rm mix}
(B_d\to\pi^+\pi^-)$ plane,\cite{FlMa2} which is shown in 
Fig.~\ref{fig:AdAmpipi} for the most recent $B$-factory data. We observe 
that the experimental averages (\ref{Bpipi-CP-averages}) and 
(\ref{Bpipi-CP-averages2}), represented by 
the crosses, overlap nicely with the SM region for $\phi_d=47^\circ$, 
and point towards $\gamma\sim60^\circ$. In this case, not only $\gamma$ 
would be in accordance with the results of the ``CKM fits'' 
(\ref{UT-Fit-ranges}), but also $\phi_d$. On the other hand, for 
$\phi_d=133^\circ$, the experimental values favour $\gamma\sim120^\circ$, 
and have essentially {\it no} overlap with the SM region. At first sight,
this may look puzzling. However, since the $\phi_d=133^\circ$ solution 
would definitely require NP contributions to $B^0_d$--$\overline{B^0_d}$ 
mixing, we may no longer use the SM interpretation of $\Delta M_d$ in 
this case to fix the UT side $R_t$, which is a crucial ingredient for the 
$\gamma$ range in (\ref{UT-Fit-ranges}). Consequently, if we choose
$\phi_d=133^\circ$, $\gamma$ may well be larger than $90^\circ$. As we 
have alread noted, the $B\to\pi K$ data seem to favour such values; a 
similar feature is also suggested by the small $B_d\to\pi^+\pi^-$ 
rate.\cite{FlMa2} Interestingly, the measured branching ratio for the 
rare kaon decay $K^+\to\pi^+\nu\overline{\nu}$ seems to point towards 
$\gamma>90^\circ$ as well,\cite{dAI} thereby also favouring the 
unconventional solution of $\phi_d=133^\circ$.\cite{FIM} Further 
valuable information on this exciting possibility can be obtained from 
the rare decays $B_{s,d}\to\mu^+\mu^-$. 

\begin{figure}[t]
$$\hspace*{-1.cm}
\epsfysize=0.19\textheight
\epsfxsize=0.29\textheight
\epsffile{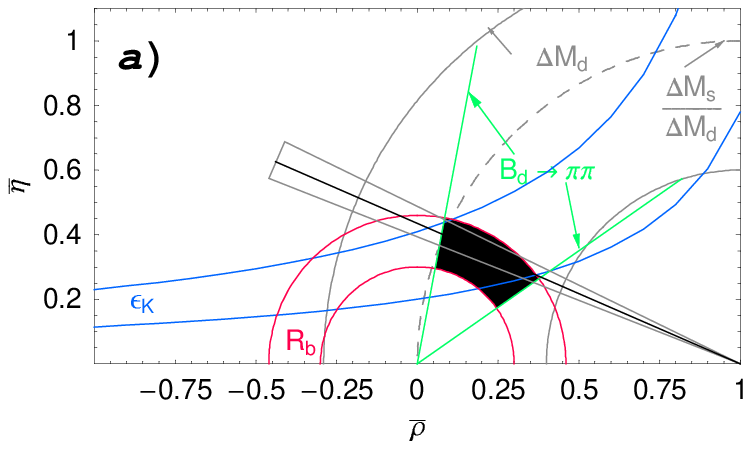} \hspace*{0.3cm}
\epsfysize=0.19\textheight
\epsfxsize=0.29\textheight
\epsffile{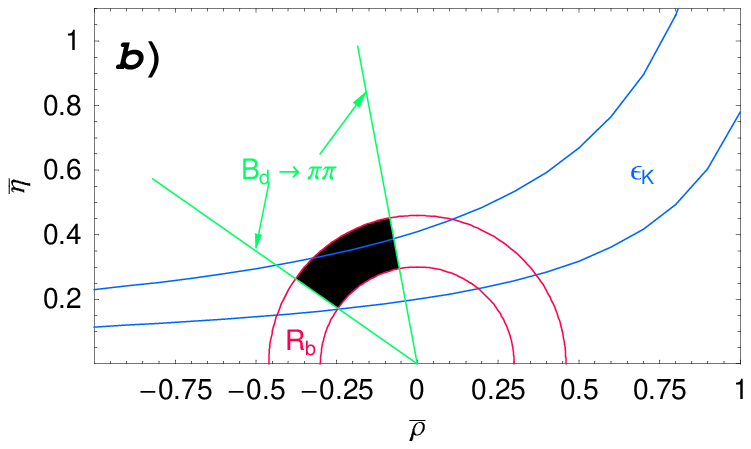}
$$
\caption[]{The allowed regions for the UT fixed through $R_b$ and CP 
violation in $B_d\to\pi^+\pi^-$, as described in the text: (a) and (b) 
correspond to $\phi_d=47^\circ$ and $\phi_d=133^\circ$, respectively 
($H=7.5$).}\label{fig:rho-eta-Bpipi}
\end{figure}

We could straightforwardly deal with this picture in a scenario for 
physics beyond the SM, where we have large NP contributions to 
$B^0_d$--$\overline{B^0_d}$ mixing, but not to the $\Delta B=1$ and 
$\Delta S=1$ decay processes. Such NP was already considered 
several years ago,\cite{GNW} and can be motivated by generic arguments 
and within supersymmetry.\cite{FIM} Since the determination of $R_b$ 
through semileptonic tree decays is in general very robust under NP effects
and would not be affected either in this particular scenario, we may 
complement $R_b$ with the range for $\gamma$ extracted from our 
$B_d\to\pi^+\pi^-$ analysis, allowing us to fix the apex of the UT in 
the $\overline{\rho}$--$\overline{\eta}$ plane. The results of this exercise 
are summarized in Fig.~\ref{fig:rho-eta-Bpipi}, following Ref.\ \refcite{FIM}, 
where also numerical values for $\alpha$, $\beta$ and $\gamma$ are given and 
a detailed discussion of the theoretical uncertainties can be found. 
Note that the SM contours implied by $\Delta M_d$, which are included in 
Fig.~\ref{fig:rho-eta-Bpipi} (a) to guide the eye, are absent in (b), 
since $B^0_d$--$\overline{B^0_d}$ mixing would there receive NP 
contributions. In this case, also we may no longer simply represent 
$\phi_d$ by a straight line, as the one in Fig.~\ref{fig:rho-eta-Bpipi} 
(a), which corresponds to $\phi_d\stackrel{{\rm SM}}{=}2\beta$, 
since we would now have $\phi_d=2\beta+\phi_d^{\rm NP}$, 
with $\phi_d^{\rm NP}\not=0^\circ$. However, we may easily read off the 
``correct'' value of $\beta$ from the black region in 
Fig.~\ref{fig:rho-eta-Bpipi} (b).\cite{FIM} Interestingly, both black
regions in Fig.~\ref{fig:rho-eta-Bpipi} (a) and (b) are consistent 
with the SM $\varepsilon_K$ hyperbola.

Because of the unsatisfactory status of the measured CP-violating 
$B_d\to\pi^+\pi^-$ observables, we may not yet draw definite 
conclusions from this analysis, although it illustrates
nicely how the corresponding strategy works. However, the experimental 
picture will improve significantly in the future, thereby providing 
more stringent constraints on $\gamma$ and the 
apex of the UT. Another milestone in this programme is the measurement 
of the CP-averaged $B_s\to K^+K^-$ branching ratio at run II of 
the Tevatron, which will allow a much better determination of $H$ that 
no longer relies on dynamical assumptions. Finally, if also the direct 
and mixing-induced CP asymmetries of $B_s\to K^+K^-$ are measured, we may
determine $\gamma$ through a minimal $U$-spin input, as
discussed above. After important steps by the CDF collaboration, LHCb and 
BTeV should be able to fully exploit the rich physics potential of the 
$B_d\to\pi^+\pi^-$, $B_s\to K^+K^-$ system. There are several 
other promising $B_s$ decays, which we shall address in the discussion of
the following section.

\boldmath
\section{New, Theoretically Clean Strategies to Extract 
$\gamma$}\label{sec:gam-clean}
\unboldmath
As far as theoretically clean determinations of $\gamma$ are
concerned, pure ``tree'' decays play the key r\^ole. In this context,
we may distinguish between the following two cases: $B$ decays exhibiting
interference effects, which are induced by subsequent $D^0,\overline{D^0}\to 
f_D$ transitions,\cite{gw}$^{\mbox{--}}$\cite{FW,APS}$^{\mbox{--}}$\cite{GGSZ} 
and channels, where neutral $B^0_q$ and $\overline{B^0_q}$ mesons may decay 
into the same final state $f$, so that we have interference effects between 
$B^0_q$--$\overline{B^0_q}$ mixing and decay processes;\cite{BdDpi,BsDsK} 
prominent examples are $B^\pm\to D K^\pm$, $B_c^\pm\to D D_s^\pm$, ...\ 
and $B_d\to D^{(\ast)\pm}\pi^\mp$, $B_s\to D_s^{(\ast)\pm} K^\mp$, ...\ 
modes, respectively. Let us focus here on recently proposed new 
strategies.\cite{RF-gam-eff-03,RF-gam-det-03,RF-gam-ca}

\boldmath
\subsection{$B_d\to D K_{\rm S(L)}$ and $B_s\to D\eta^{(')}, 
D\phi$, ...}\label{ssec:BqDfr}
\unboldmath
Colour-suppressed $B^0_d\to D^0 K_{\rm S}$ decays and similar modes 
provide interesting tools to explore CP violation.\cite{GroLo,KayLo} 
In the following, we shall consider general transitions of the kind 
$B^0_q\to D^0f_r$, where $r\in\{s,d\}$ distinguishes between 
$b\to Ds$ and $b\to D d$ processes.\cite{RF-gam-eff-03,RF-gam-det-03} If we 
require $({\cal CP})|f_r\rangle=\eta_{\rm CP}^{f_r}|f_r\rangle$, $B^0_q$ and
$\overline{B^0_q}$ mesons may both decay into $D^0f_r$, thereby leading to 
interference between $B^0_q$--$\overline{B^0_q}$ mixing and decay processes, 
which involve $\phi_q+\gamma$. In the case of $r=s$, corresponding to 
$B_d\to D K_{\rm S(L)}$, $B_s\to D\eta^{(')}, D\phi$, ..., these 
interference effects are governed by a hadronic parameter 
$x_{f_s}e^{i\delta_{f_s}}\propto R_b\approx0.4$, and are hence favourably 
large. On the other hand, for $r=d$, which describes 
$B_s\to DK_{\rm S(L)}$, $B_d\to D\pi^0, D\rho^0$ ...\ modes, the interference 
effects are tiny because of $x_{f_d}e^{i\delta_{f_d}}\propto -\lambda^2R_b
\approx -0.02$. Let us first focus on the $r=s$ case. 

If we consider $B_q\to D_\pm f_s$ modes, where
$({\cal CP})|D_\pm\rangle=\pm|D_\pm\rangle$, additional interference between 
$B_q^0\to D^0 f_s$ and $B_q^0\to \overline{D^0} f_s$ arises at the decay 
level, involving $\gamma$. The most straightforward observable 
we may measure is the ``untagged'' rate
\begin{eqnarray}
\lefteqn{\langle\Gamma(B_q(t)\to D_\pm f_s)\rangle\equiv
\Gamma(B^0_q(t)\to D_\pm f_s)+
\Gamma(\overline{B^0_q}(t)\to D_\pm f_s)}\\
&&\stackrel{\Delta\Gamma_q=0}{=}
\left[\Gamma(B^0_q\to D_\pm f_s)+
\Gamma(\overline{B^0_q}\to D_\pm f_s)\right]e^{-\Gamma_qt}\equiv 
\langle\Gamma(B_q\to D_\pm f_s)\rangle e^{-\Gamma_qt},\nonumber
\end{eqnarray}
which allows us to determine the following ``untagged'' rate asymmetry:
\begin{equation}
\Gamma_{+-}^{f_s}\equiv
\frac{\langle\Gamma(B_q\to D_+ f_s)\rangle-\langle
\Gamma(B_q\to D_- f_s)\rangle}{\langle\Gamma(B_q\to D_+ f_s)\rangle
+\langle\Gamma(B_q\to D_- f_s)\rangle},
\end{equation}
already providing very interesting information about 
$\gamma$:\cite{RF-gam-eff-03} first, we may derive 
\begin{equation}
|\cos\gamma|\geq |\Gamma_{+-}^{f_s}|, 
\end{equation}
which implies bounds on $\gamma$. Second, taking into account that 
factorization suggests $\cos\delta_{f_s}>0$,\cite{RF-gam-det-03} we obtain 
\begin{equation}\label{sgn-cos-gam}
\mbox{sgn}(\cos\gamma)=\mbox{sgn}(\Gamma_{+-}^{f_s}),
\end{equation} 
allowing us to decide whether $\gamma$ is smaller or larger than 
$90^\circ$.

If we measure also the mixing-induced observables 
$S_\pm^{f_s}\equiv {\cal A}_{\rm CP}^{\rm mix}(B_q\to D_\pm f_s)$, 
we may determine $\gamma$. To this end, it is convenient to introduce
the quantities
\begin{equation}
\langle S_{f_s}\rangle_\pm\equiv\frac{S_+^{f_s}\pm S_-^{f_s}}{2}.
\end{equation}
Expressing the $\langle S_{f_s}\rangle_\pm$ in terms of the $B_q\to D_\pm f_s$ 
decay parameters gives rather complicated formulae. However, 
complementing the $\langle S_{f_s}\rangle_\pm$ with $\Gamma_{+-}^{f_s}$ 
yields 
\begin{equation}\label{key-rel}
\tan\gamma\cos\phi_q=
\left[\frac{\eta_{f_s} \langle S_{f_s}
\rangle_+}{\Gamma_{+-}^{f_s}}\right]+\left[\eta_{f_s}\langle S_{f_s}\rangle_--
\sin\phi_q\right],
\end{equation}
where $\eta_{f_s}\equiv(-1)^L\eta_{\rm CP}^{f_s}$, with $L$ denoting the 
$Df_s$ angular momentum.\cite{RF-gam-eff-03} Using this simple -- 
but {\it exact} -- relation, we obtain the twofold solution 
$\gamma=\gamma_1\lor\gamma_2$, with $\gamma_1\in[0^\circ,180^\circ]$ and 
$\gamma_2=\gamma_1+180^\circ$. Since $\cos\gamma_1$ and $\cos\gamma_2$
have opposite signs, (\ref{sgn-cos-gam}) allows us to fix $\gamma$ 
{\it unambiguously}. Another advantage of (\ref{key-rel}) is that 
$\langle S_{f_s}\rangle_+$ and $\Gamma_{+-}^{f_s}$ are both proportional 
to $x_{f_s}\approx0.4$, so that the first term in square brackets is 
of ${\cal O}(1)$, whereas the second one is of ${\cal O}(x_{f_s}^2)$, 
hence playing a minor r\^ole. In order to extract 
$\gamma$, we may also employ $D$ decays into CP non-eigenstates 
$f_{\rm NE}$, where we have to deal with complications originating from 
$D^0,\overline{D^0}\to f_{\rm NE}$ interference effects.\cite{KayLo}
Also in this case, $\Gamma_{+-}^{f_s}$ is a very powerful ingredient, 
offering an efficient, {\it analytical} strategy to include these 
interference effects in the extraction of $\gamma$.\cite{RF-gam-det-03}

Let us briefly come back to the $r=d$ case, corresponding to 
$B_s\to DK_{\rm S(L)}$, $B_d\to D\pi^0, D\rho^0$ ...\ decays, which 
can be described through the same formulae as their $r=s$ counterparts. 
Since the relevant interference effects are governed by
$x_{f_d}\approx -0.02$, these channels are not as attractive for the 
extraction of $\gamma$ as the $r=s$ modes. On the other 
hand, the relation
\begin{equation}
\eta_{f_d}\langle S_{f_d}\rangle_-=
\sin\phi_q + {\cal O}(x_{f_d}^2)
=\sin\phi_q + {\cal O}(4\times 10^{-4})
\end{equation}
offers very interesting determinations of $\sin\phi_q$.\cite{RF-gam-eff-03} 
Following this avenue, there are no penguin uncertainties, and the theoretical 
accuracy is one order of magnitude better than in the ``conventional'' 
$B_d\to J/\psi K_{\rm S}$, $B_s\to J/\psi \phi$ strategies. In particular,
$\phi_s^{\rm SM}=-2\lambda^2\eta$ could, in principle, be determined with 
a theoretical uncertainty of only ${\cal O}(1\%)$, which should be
compared with the discussion given in Subsection~\ref{ssec:Bspsiphi}.
The $\overline{B^0_d}\to D^0\pi^0$ mode has already been seen at the 
$B$ factories, with branching ratios at the 
$3\times 10^{-4}$ level.\cite{Bbar-D0pi0} Recently, the Belle collaboration
has also announced the observation of $\overline{B^0_d}\to D^0 
\overline{K^0}$, with the branching ratio 
$(5.0^{+1.3}_{-1.2}\pm0.6)\times10^{-5}$.\cite{Belle-BdDK-obs}

\boldmath
\subsection{$B_s\to D_s^{(\ast)\pm} K^\mp, ...$ and 
$B_d\to D^{(\ast)\pm} \pi^\mp, ...$}
\unboldmath
Let us now turn to the colour-allowed counterparts of the 
$B_q\to D f_q$ modes discussed in Subsection \ref{ssec:BqDfr}, which 
we may write generically as $B_q\to D_q \overline{u}_q$.\cite{RF-gam-ca}
The characteristic feature of these transitions is that both a $B^0_q$
and a $\overline{B^0_q}$ meson may decay into $D_q \overline{u}_q$, 
thereby leading to interference between $B^0_q$--$\overline{B^0_q}$ 
mixing and decay processes, which involve the weak phase $\phi_q+\gamma$.
In the case of $q=s$, which corresponds to $D_s\in\{D_s^+, D_s^{\ast+}, ...\}$ 
and $u_s\in\{K^+, K^{\ast+}, ...\}$, these interference effects are
governed by a hadronic parameter $x_s e^{i\delta_s}\propto R_b\approx0.4$,
and hence are large. On the other hand, in the case of $q=d$, corresponding 
to $D_d\in\{D^+, D^{\ast+}, ...\}$ and $u_d\in\{\pi^+, \rho^+, ...\}$, 
they are described by $x_d e^{i\delta_d}\propto -\lambda^2R_b\approx-0.02$, 
and hence are tiny. In the following, we shall only consider 
$B_q\to D_q \overline{u}_q$ modes, where at least one of the 
$D_q$, $\overline{u}_q$ states is a pseudoscalar meson; otherwise a 
complicated angular analysis has to be performed.\cite{FD2,LSS,GPW}
 
It is well known that such decays allow a determination of
$\phi_q+\gamma$, where the ``conventional'' approach works
as follows:\cite{BdDpi,BsDsK} if we measure the observables 
$C(B_q\to D_q\overline{u}_q)\equiv C_q$ 
and $C(B_q\to \overline{D}_q u_q)\equiv \overline{C}_q$ provided by the
$\cos(\Delta M_qt)$ pieces of the time-dependent rate asymmetries, 
we may determine $x_q$ from terms entering at the $x_q^2$ 
level. In the case of $q=s$, we have $x_s={\cal O}(R_b)$, 
implying $x_s^2={\cal O}(0.16)$, so that this may actually be possible, 
though challenging. On the other hand, 
$x_d={\cal O}(-\lambda^2R_b)$ is doubly Cabibbo-suppressed. 
Although it should be possible to resolve terms of ${\cal O}(x_d)$, 
this will be impossible for the vanishingly small $x_d^2={\cal O}(0.0004)$ 
terms, so that other approaches to fix $x_d$ are required.\cite{BdDpi}
In order to extract $\phi_q+\gamma$, we must measure the 
mixing-induced observables $S(B_q\to D_q\overline{u}_q)\equiv S_q$ and 
$S(B_q\to \overline{D}_q u_q)\equiv \overline{S}_q$ associated with the
$\sin(\Delta M_qt)$ terms of the time-dependent rate 
asymmetries, where it is convenient to introduce
\begin{equation}\label{Savpm}
\langle S_q\rangle_\pm\equiv
\frac{\overline{S}_q\pm S_q}{2}.
\end{equation}
If we assume that the hadronic parameter $x_q$ is known, we may consider
\begin{eqnarray}
s_+&\equiv& (-1)^L
\left[\frac{1+x_q^2}{2 x_q}\right]\langle S_q\rangle_+
=+\cos\delta_q\sin(\phi_q+\gamma)\\
s_-&\equiv&(-1)^L
\left[\frac{1+x_q^2}{2x_q}\right]\langle S_q\rangle_-
=-\sin\delta_q\cos(\phi_q+\gamma),
\end{eqnarray}
yielding
\begin{equation}\label{conv-extr}
\sin^2(\phi_q+\gamma)=\frac{1}{2}\left[(1+s_+^2-s_-^2)\pm
\sqrt{(1+s_+^2-s_-^2)^2-4s_+^2}\right],
\end{equation}
which implies an eightfold solution for $\phi_q+\gamma$; assuming
$\mbox{sgn}(\cos\delta_q)>0$, as suggested by factorization, a fourfold 
discrete ambiguity emerges. This assumption allows us also to extract 
the sign of $\sin(\phi_q+\gamma)$ from $\langle S_q\rangle_+$. To this
end, the factor $(-1)^L$, where $L$ is the $D_q\overline{u}_q$ angular 
momentum, has to be properly taken into account.\cite{RF-gam-ca} 
This is crucial for the extraction of the sign of 
$\sin(\phi_d+\gamma)$ from $B_d\to D^{\ast\pm}\pi^\mp$ 
modes, allowing us to distinguish between the two solutions shown in
Fig.\ \ref{fig:rho-eta-Bpipi}.

Let us now discuss new approaches to deal with $B_q\to D_q \overline{u}_q$
modes, following Ref.\ \refcite{RF-gam-ca}. If $\Delta\Gamma_s$ is 
sizeable, the ``untagged'' rates 
\begin{eqnarray}
\lefteqn{\langle\Gamma(B_q(t)\to D_q\overline{u}_q)\rangle=
\langle\Gamma(B_q\to D_q\overline{u}_q)\rangle}\nonumber\\
&&\times\left[\cosh(\Delta\Gamma_qt/2)-{\cal A}_{\rm \Delta\Gamma}
(B_q\to D_q\overline{u}_q)\,\sinh(\Delta\Gamma_qt/2)\right]
e^{-\Gamma_qt}\label{untagged}
\end{eqnarray}
provide observables ${\cal A}_{\rm \Delta\Gamma}(B_s\to D_s\overline{u}_s)
\equiv {\cal A}_{\rm \Delta\Gamma_s}$ and 
${\cal A}_{\rm \Delta\Gamma}(B_s\to \overline{D}_s u_s)\equiv 
\overline{{\cal A}}_{\rm \Delta\Gamma_s}$, which yield 
\begin{equation}\label{untagged-extr}
\tan(\phi_s+\gamma)=
-\left[\frac{\langle S_s\rangle_+}{\langle{\cal A}_{\rm \Delta\Gamma_s}
\rangle_+}\right]
=+\left[\frac{\langle{\cal A}_{\rm \Delta\Gamma_s}
\rangle_-}{\langle S_s\rangle_-}\right],
\end{equation}
where $\langle{\cal A}_{\rm \Delta\Gamma_s}\rangle_\pm$ is defined
in analogy to (\ref{Savpm}). These relations allow an 
{\it unambiguous} determination of $\phi_s+\gamma$, if we employ again
$\mbox{sgn}(\cos\delta_q)>0$. Another important
advantage of (\ref{untagged-extr}) is that we have {\it not} to rely on 
${\cal O}(x_s^2)$ terms, as $\langle S_s\rangle_\pm$ and 
$\langle {\cal A}_{\rm \Delta\Gamma_s}\rangle_\pm$ are proportional to $x_s$. 
On the other hand, we need a sizeable value of $\Delta\Gamma_s$. Measurements 
of untagged rates are also very useful in the case of vanishingly small 
$\Delta\Gamma_q$, since the ``unevolved'' untagged rates 
in (\ref{untagged}) offer various interesting strategies to determine 
$x_q$ from the ratio of $\langle\Gamma(B_q\to D_q\overline{u}_q)\rangle+
\langle\Gamma(B_q\to \overline{D}_q u_q)\rangle$ and CP-averaged rates of
appropriate $B^\pm$ or flavour-specific $B_q$ decays.

If we keep the hadronic parameter $x_q$ and the associated strong phase
$\delta_q$ as ``unknown'', free parameters in the expressions for the
$\langle S_q\rangle_\pm$, we obtain 
\begin{equation}
|\sin(\phi_q+\gamma)|\geq|\langle S_q\rangle_+|, \quad
|\cos(\phi_q+\gamma)|\geq|\langle S_q\rangle_-|,
\end{equation}
which can straightforwardly be converted into bounds on $\phi_q+\gamma$. 
If $x_q$ is known, stronger constraints are implied by 
\begin{equation}\label{bounds}
|\sin(\phi_q+\gamma)|\geq|s_+|, \quad
|\cos(\phi_q+\gamma)|\geq|s_-|.
\end{equation}
Once $s_+$ and $s_-$ are known, we may of course determine
$\phi_q+\gamma$ through the ``conventional'' approach, using 
(\ref{conv-extr}). However, the bounds following from (\ref{bounds})
provide essentially the same information and are much simpler to 
implement. Moreover, as discussed in detail in Ref.\ \refcite{RF-gam-ca}
for several examples corresponding to the SM, the bounds following from
$B_s$ and $B_d$ modes may be highly complementary, thereby providing
particularly narrow, theoretically clean ranges for $\gamma$. 

If we look at the corresponding topologies, we observe that
$B_s^0\to D_s^{(\ast)+}K^-$ and $B_d^0\to D^{(\ast)+}\pi^-$ are related 
to each other through an interchange of all down and strange quarks. 
Consequently, the $U$-spin symmetry implies $a_s=a_d$ and $\delta_s=\delta_d$,
where $a_s=x_s/R_b$ and $a_d=-x_d/(\lambda^2 R_b)$ are the ratios of
hadronic matrix elements entering $x_s$ and $x_d$, respectively. There
are various possibilities to implement these relations.\cite{RF-gam-ca}
For example, we may assume that $a_s=a_d$ {\it and} $\delta_s=\delta_d$,
yielding
\begin{equation}
\tan\gamma=-\left[\frac{\sin\phi_d-S
\sin\phi_s}{\cos\phi_d-S\cos\phi_s}
\right]\stackrel{\phi_s=0^\circ}{=}
-\left[\frac{\sin\phi_d}{\cos\phi_d-S}\right],
\end{equation}
where 
\begin{equation}
S=-R\left[\frac{\langle S_d\rangle_+}{\langle S_s\rangle_+}\right],
\end{equation}
with
\begin{equation}
R= \left(\frac{1-\lambda^2}{\lambda^2}\right)
\left[\frac{1}{1+x_s^2}\right];
\end{equation}
$R$ can be fixed through untagged $B_s$ rates with the help of 
\begin{equation}
R=\left(\frac{f_K}{f_\pi}\right)^2 
\left[\frac{\Gamma(\overline{B^0_s}\to D_s^{(\ast)+}\pi^-)+
\Gamma(B^0_s\to D_s^{(\ast)-}\pi^+)}{\langle\Gamma(B_s\to D_s^{(\ast)+}K^-)
\rangle+\langle\Gamma(B_s\to D_s^{(\ast)-}K^+)\rangle}\right].
\end{equation}
Alternatively, we may {\it only} assume that $\delta_s=\delta_d$ {\it or} 
that $a_s=a_d$, as discussed in detail in Ref.\ \refcite{RF-gam-ca}. 
Apart from features related to multiple discrete ambiguities, the 
most important advantage in comparison with the ``conventional'' approach 
is that the experimental resolution of the $x_q^2$ terms is not required. In 
particular, $x_d$ does {\it not} have to be fixed, and $x_s$ may only enter 
through a $1+x_s^2$ correction, which can straightforwardly be determined 
through 
untagged $B_s$ rate measurements. In the most refined implementation of this 
strategy, the measurement of $x_d/x_s$ would {\it only} be interesting for 
the inclusion of $U$-spin-breaking effects in $a_d/a_s$. Moreover, we may 
obtain interesting insights into hadron dynamics and $U$-spin-breaking 
effects.

\section{Conclusions and Outlook}\label{sec:concl}
Thanks to the ``gold-plated'' mode $B_d\to J/\psi K_{\rm S}$ and similar
channels, CP violation is now a well established phenomenon in the 
$B$-meson system. Although the present world average 
$\sin\phi_d=0.734\pm0.054$ agrees well with the SM, we obtain the 
twofold solution $\phi_d\sim 47^\circ \lor 133^\circ$, where the latter 
leaves us with the exciting possibility of having large NP contributions 
to $B^0_d$--$\overline{B^0_d}$ mixing. The $B$ factories allow us now 
to confront many strategies to explore CP violation with the first data, 
yielding the following present picture: the SM relation 
${\cal A}_{\rm CP}^{\rm mix}(B_d\to \phi K_{\rm S})=
{\cal A}_{\rm CP}^{\rm mix}(B_d\to J/\psi K_{\rm S})$ may not be satisfied,
$B\to\pi K$ data point towards $\gamma\gsim90^\circ$ and a ``puzzling'' 
picture for strong phases, CP violation in $B_d\to\pi^+\pi^-$ could
accommodate $\gamma>90^\circ$ for $\phi_d=133^\circ$. The experimental 
uncertainties do not yet allow us to draw definite conclusions,
but the situation will significantly improve in the future. 

The physics potential of the $e^+e^-$ $B$ factories operating at the
$\Upsilon(4S)$ resonance is nicely complemented by $B$-decay studies
at hadron colliders, run II of the Tevatron and the LHC, providing 
in particular 
access to the $B_s$-meson system.  Already a measurement of the mass 
difference $\Delta M_s$ would be a very important achievement, since
$\Delta M_s/\Delta M_d$ provides a particularly interesting determination 
of the side $R_t$ of the UT. However, the most exciting aspects are 
related to the exploration of CP violation, which is offered by several 
promising $B_s$ modes. Here another ``gold-plated'' mode is given by
$B_s\to J/\psi\phi$, allowing us to check whether $\phi_s$ is actually 
negligibly small, as expected in the SM, or whether this phase is 
enhanced through NP contributions to $B^0_s$--$\overline{B^0_s}$ mixing; 
also in the latter case, we could extract $\phi_s$ through mixing-induced 
CP violation in $B_s\to J/\psi\phi$. As we have seen, another very 
interesting channel is the penguin-dominated decay $B_s\to K^+K^-$, 
which complements $B_d\to\pi^+\pi^-$, thereby offering a promising 
determination of $\gamma$. Moreover, there are theoretically clean 
strategies to extract $\gamma$, where pure ``tree'' decays of $B_s$ 
mesons play a key r\^ole. Here it is also of great advantage to 
complement $B_s$ with $B_d$ modes, considering, for example, the 
$B_s\to D_s^{(\ast)\pm}K^\mp$, $B_d\to D^{(\ast)\pm}\pi^\mp$ system. 
It will be very exciting to see whether discrepancies between the 
$B_s\to K^+K^-$, $B_d\to\pi^+\pi^-$ and $B_s\to D_s^{(\ast)\pm}K^\mp$, 
$B_d\to D^{(\ast)\pm}\pi^\mp$ results for $\gamma$ will emerge. 
Detailed feasibility studies of the new strategies discussed above 
are strongly encouraged.

\end{document}